\documentclass[aps,prd,twocolumn,groupedaddress,showpacs,preprintnumbers,draft]
{revtex4}
\usepackage{epsf}
\usepackage{bm}
\usepackage{amsfonts}

\begin{document}
\preprint{BROWN-HET-1436}
\preprint{MCGILL-01-05}
\preprint{NORDITA-07-05}
\def\Box{\nabla^2}
\def\be{\begin{equation}}
\def\ee{\end{equation}}
\def\bea{\begin{eqnarray}}
\def\eea{\end{eqnarray}}


\newcommand{\mx}{\mbox}
\newcommand{\mt}{\mathtt}
\newcommand{\mtf}{\mathtt{f}}
\newcommand{\mts}{\mathtt{s}}
\newcommand{\mtr}{\mathtt{r}}
\newcommand{\mtb}{\mathtt{b}}
\newcommand{\mtm}{\mathtt{m}}
\newcommand{\mtd}{\mathtt{d}}
\newcommand{\mtc}{\mathtt{c}}
\newcommand{\mtw}{\mathtt{w}}
\newcommand{\p}{\partial}
\newcommand{\st}{\stackrel}
\newcommand{\al}{\alpha}
\newcommand{\bb}{\beta}
\newcommand{\ga}{\gamma}
\newcommand{\te}{\theta}
\newcommand{\de}{\delta}
\newcommand{\et}{\eta}
\newcommand{\ze}{\zeta}
\newcommand{\s}{\sigma}
\newcommand{\e}{\epsilon}
\newcommand{\om}{\omega}
\newcommand{\Om}{\Omega}
\newcommand{\la}{\lambda}
\newcommand{\La}{\Lambda}
\newcommand{\ti}{\widetilde}
\newcommand{\hn}{\widehat{\nabla}}
\newcommand{\ah}{\widehat{a}}
\newcommand{\bh}{\widehat{b}}
\newcommand{\ch}{\widehat{c}}
\newcommand{\ddh}{\widehat{d}}
\newcommand{\eh}{\widehat{e}}
\newcommand{\gh}{\widehat{g}}
\newcommand{\ph}{\widehat{p}}
\newcommand{\qh}{\widehat{q}}
\newcommand{\ih}{\widehat{i}}
\newcommand{\jh}{\widehat{j}}
\newcommand{\kh}{\widehat{k}}
\newcommand{\lh}{\widehat{l}}
\newcommand{\mh}{\widehat{m}}
\newcommand{\nh}{\widehat{n}}
\newcommand{\Dh}{\widehat{D}}
\newcommand{\sto}{\st{\circ}}
\newcommand{\as}{\st{\circ}{a}}
\newcommand{\bs}{\st{\circ}{b}}
\newcommand{\cs}{\st{\circ}{c}}
\newcommand{\ds}{\st{\circ}{d}}
\newcommand{\es}{\st{\circ}{e}}
\newcommand{\gs}{\st{\circ}{g}}
\newcommand{\ms}{\st{\circ}{m}}
\newcommand{\ns}{\st{\circ}{n}}
\newcommand{\ps}{\st{\circ}{p}}
\newcommand{\Ds}{\st{\circ}{D}}
\newcommand{\2}{\frac{1}{2}}
\newcommand{\3}{\textstyle{1\over 3}}
\newcommand{\4}{\frac{1}{4}}
\newcommand{\8}{\textstyle{1\over 8}}
\newcommand{\6}{\textstyle{1\over 16}}
\newcommand{\ra}{\rightarrow}
\newcommand{\lra}{\longrightarrow}
\newcommand{\Ra}{\Rightarrow}
\newcommand{\im}{\Longleftrightarrow}
\newcommand{\hs}{\hspace{5mm}}
\newcommand{\vs}{\vspace{5mm}\\}

\title{Coupled  Inflation and Brane Gases}
\author{Tirthabir Biswas $^{1)}$\footnote{tirtho@hep.physics.mcgill.ca},
Robert Brandenberger $^{1,2)}$\footnote{rhb@hep.physics.mcgill.ca},
Damien A.~Easson $^{3)}$\footnote{easson@physics.syr.edu},
and Anupam Mazumdar $^{4,5)}$\footnote{anupamm@nordita.dk}}
\affiliation{1) Department of Physics, McGill University, Montr\'eal, QC,
H3A 2T8, CANADA}
\affiliation{2) Department of Physics, Brown University, Providence, RI 02912,
USA}
\affiliation{3) Department of Physics, Syracuse University, Syracuse, NY 13244,
USA}
\affiliation{4) NORDITA, Blegdamsvej 17, Copenhagen 2100, DENMARK.}
\affiliation{5) The Niels Bohr Institute, Blegdamsvej-17, Copenhagen, DENMARK}

\begin{abstract}
We study an effective four-dimensional theory with an action with two scalar
fields minimally coupled to gravity, and with a matter action which
couples to the two scalar fields via an overall field-dependent
coefficient in the action. Such a theory could arise from a dimensional
reduction of supergravity coupled to a gas of branes winding the compactified
dimensions. We show the existence of solutions corresponding to power-law 
inflation. The graceful exit from inflation can be obtained by postulating
the decay of the branes, as would occur if the branes are unstable in the
vacuum and stabilized at high densities by plasma effects. This construction
provides an avenue for connecting string gas cosmology and the late-time
universe. 
\end{abstract}
\maketitle

\section{Introduction}

String gas cosmology, originally discussed in \cite{BV,TV} (see also
\cite{Perlt}) and later reconstructed in the context of branes as
fundamental degrees of freedom in string theory~\cite{ABE,BEK}, provides a
mechanism for constructing a nonsingular cosmology and a mechanism
which confines all but three spatial dimensions to be microscopic in
size. String winding modes play a key role in this scenario: they are
unable to annihilate in more than three spatial dimensions \cite{BV}
because the probability that two string world sheets intersect is of
measure zero (see \cite{caveats} for some recent caveats regarding
this scenario).  A key assumption in string gas cosmology is that all
spatial dimensions start out at string scale, and that the universe is
hot (and brane degrees of freedom excited). However, in order to
connect this scenario to our current universe, a mechanism is required
which expands the volume of our observed three spatial dimensions to
be large enough to contain the currently observed Hubble volume. This
is one aspect of the ``flatness problem'' of Standard Big Bang
cosmology (see e.g. \cite{Guth} for a discussion), and a period of
cosmological inflation after the winding modes in our three spatial
dimensions have disappeared is the only currently known solution.

Thus, an outstanding challenge for string gas cosmology has been to
provide a stringy realization of inflation \footnote{It has recently
been shown that a conventional four-dimensional realization of
inflation using regular scalar fields in the four-dimensional
effective field theory is not consistent with the late-time
stabilization of the extra dimensions \cite{Patil}.}. One suggestion
was recently made in \cite{BEM} and made use of a gas of co-dimension
one branes which provide a period of power-law inflation, with an
equation of state $\om = -2/3$ (where $\om = p / \rho$,  and $p$ and $\rho$
are the pressure and energy density, respectively), which is however
incompatible with inflation being the source of the cosmic microwave
anisotropies (see e.g. \cite{WMAP}). The graceful exit from inflation
was achieved by considering the branes to be unstable in the vacuum,
and stabilized by plasma effects in a hot gas, in analogy to how
embedded defects in field theory can be stabilized in the plasma of
the early universe \cite{Nag,Carter}.

In this paper we consider a more natural realization of inflation in
the context of string gas cosmology. We study the period of cosmology
after our three dimensions have shredded their winding modes (as
discussed in detail in \cite{BEK}), but the other dimensions are still
confined by strings and branes which wind around them. We consider the
four-dimensional effective action, and focus in particular on two
scalar fields in this action, one corresponding to a modulus of the
higher-dimensional theory, the other corresponding to the radion. We
assume an exponential potential for the modulus field and consider
brane string matter (treated as a hydrodynamical fluid) whose overall
action has a prefactor which depends exponentially on the two scalar
fields. We then demonstrate that it is possible to obtain solutions of
power-law inflation with an effective equation of state more negative
than $\om = -2/3$. The coupling of matter to the scalar fields
required for acceleration arises automatically if we consider branes
winding around the compact space as our matter. To obtain a graceful
exit from inflation, we again require the branes to be unstable, as in
\cite{BEM}.  Compared to \cite{BEM}, our mechanism may provide an
inflationary phase with a smaller value of $\om$, and it does not
require branes which are extended in our large spatial dimensions. Our
mechanism relies of the analysis of \cite{BM} which we now review.

\section{Coupled Inflation}

In \cite{BM} a mechanism to realize an accelerated phase of expansion
was proposed which relies on a rolling scalar field being coupled
(roughly with gravitational strength) to some form of matter or
radiation.  Although the main motivation in \cite{BM} came from trying
to explain the late time acceleration phase that we seem to be
entering into today, in this paper we show how one can obtain
inflation with a gas of unstable branes via a similar mechanism. Since
branes in general couple to both the dilaton and the radion (the
volume of the extra dimensions), unless one invokes an additional
mechanism to stabilize one of the scalars, one has to study the
dynamics involving both the radion and the dilaton. Hence we first
review and generalize the analysis done in \cite{BM}.

{\bf Evolution Equations:} We start with an effective four dimensional 
action with two scalar fields minimally coupled to gravity
\bea
{\hat{S}} \, &=& \, \frac{M_p^2}{2} \int d^{4}x\sqrt{-g}\bigl[R - 
\p_{\mh}\phi\p^{\mh}\phi \\
&-& \p_{\mh}\psi\p^{\mh}\psi-\frac{2V_0}{M_p^2}e^{-2(\al\phi+\bb\psi)}\bigr] \, .\nonumber
\eea
where $M_p$ is the reduced Planck mass which we will set to 1 from now
on.  Such exponential potentials are common in dimensionally reduced
supergravity/M-theories in dilaton-radion systems~\cite{exppot}, or in the context
of large extra dimensions~\cite{mohapatra} and we will give some
explicit examples later. In the above, $\al,\bb$ are constants which
depend on the origin of the potential as well as on the dimensionality
of the original model. Consider now that a form of matter or radiation
also couples to the scalar fields:
\be
S_{\mt{mat}}=\int d^4x \sqrt{-g}\ti{\rho}=\int d^4x \sqrt{-g}\rho e^{2(\mu\psi+\nu\phi)}
\ee
where, as usual, $\ti{\rho}$ is the observed energy density and we
define a ``bare density'', $\rho\equiv\ti{\rho}
e^{-2(\mu\psi+\nu\phi)}$, which obeys the familiar evolution equations
involving an equation of state parameter $\om$
\be
\dot{\rho} + 3H(p+\rho) \, = \, 0 
\ee
with 
\be
p=\om\rho \Ra \rho=\rho_0\left(\frac{a}{a_0}\right)^{-3(1+\om)} \, .
\label{rho}
\ee
In the next section we will identify $\ti{\rho}$ with the energy
density of a gas of branes wrapping extra dimensions and will derive
the coupling exponents $\mu$ and $\nu$ explicitly.

Due to this coupling, the brane action not only acts as a source term
for gravity but also provides an effective potential term for the
scalar fields. The field equations for the scalars now read:
\be
\ddot{\phi}+3H\dot{\phi}=-\frac{\p V_{\mt{eff}}(\phi,\psi)}{\p \phi}
\ee
and
\be
\ddot{\psi}+3H\dot{\psi}=-\frac{\p V_{\mt{eff}}(\phi,\psi)}{\p \psi}
\ee
where
\be
V_{\mt{eff}}(\phi)=V_0e^{-2(2\al\psi+\bb \phi)}+e^{2(\mu \psi+\nu\psi)}\rho 
\, . \label{effective}
\ee
The above multi-field exponential potential has been studied in the
context of {\it assisted inflation}, see~\cite{Liddle-maz}. Similar
multi-field potentials were also used to construct inflationary models
in \cite{ACD1,ACD2}. {F}rom
here onwards it is useful to work in terms of rotated
fields~\cite{wands},
\be
\left(\begin{array}{c}
\psi'\\
\phi'
\end{array}\right)=
\left(\begin{array}{cc}
\cos\te&\sin\te\\
-\sin\te&\cos\te
\end{array}\right)
\left(\begin{array}{c}
\psi\\
\phi
\end{array}\right)
\label{linear}
\ee
where we choose
\be
\cos\te=\frac{\mu}{\sqrt{\mu^2+\nu^2}}
\ee
such that 
\be
\ti{\rho}=\rho e^{2\mu'\psi'}\ \mx{ with } \ \mu'=\sqrt{\mu^2+\nu^2}
\ee
depends only on a single field.  The exponents $\al,\bb$ transform in the 
same way as the fields:
\be
\left(\begin{array}{c}
\al'\\
\bb'
\end{array}\right)=
\left(\begin{array}{cc}
\cos\te&\sin\te\\
-\sin\te&\cos\te
\end{array}\right)
\left(\begin{array}{c}
\al\\
\bb
\end{array}\right)
\ee

Thus, the ``rotated'' field equations read as
\be
\ddot{\phi'}+3H\dot{\phi'}=2\bb' V_0e^{-2(\al'\psi'+\bb' \phi')}
\label{phi}
\ee
and
\be
\ddot{\psi'}+3H\dot{\psi'}=2\al' V_0e^{-2(\al'\psi'+\bb' \phi')}-2\mu'\rho e^{2\mu' \psi'}
\label{psi}
\ee

One can also write down
the Friedmann equation for a Robertson Walker metric:
\begin{eqnarray}
\label{Hubble}
H^2 =\3\left( \frac{\dot{\phi'}^2}{2}+\frac{\dot{\psi'}^2}{2}+V_0e^{-2(\al'\psi'+\bb' \phi')}+\rho e^{2\mu' \psi'}\right)\,
\end{eqnarray}

{\bf Inflation from a Single Field:} Before we solve
(\ref{phi}-\ref{Hubble}) in its full generality, it is insightful to
look at a special case when the dynamics reduces to that of a single
field which has been studied in detail in \cite{BM}. This happens when
\be
\frac{\bb}{\al}=\tan\te=\frac{\nu}{\mu}\Ra \bb'=0
\ee
In this case, $\phi'$ remains frozen (even if one starts with a
non-zero $\dot{\phi'}$ it is quickly Hubble damped) and $\psi'$
evolves under the influence of the effective potential
(\ref{effective}) with exponents $\al'=\sqrt{\al^2+\bb^2}$ and
$\mu'$. As discussed in \cite{BM}, provided the exponents are of
${\cal O}(1)$ and have the same sign, $\psi'$ tracks the minimum
formed between the two opposing exponentials.  Since $\rho$ redshifts,
the minimum redshifts and one can solve for the evolution of $\psi'$
and $a(t)$ to get
\bea \label{psi-evol}
e^{2\psi'} &=& \left(\frac{\al'V_0}{\mu'\rho}\right)^{1/(\mu'+\al')} \\
&=& \left(\frac{\al'V_0}{\mu'\rho_0}\right)^{1/(\mu'+\al')}\left(\frac{a}{a_0}\right)^{3(1+\om)/(\mu'+\al')} \nonumber
\eea
and
\be
\label{scalefact}
a(t)=a_0\left(\frac{t}{t_0}\right)^{(2/3\al')
\left[(\mu'+\al')/(1+\om)\right]}\,.
\label{a-evol}
\ee
Thus, the universe accelerates if
\be
\frac{\mu'}{\al'}> \2(1+3\om)
\label{acceleration}
\ee
{F}or non-relativistic dust type matter, like wrapped branes, we have
$\om=0$, so that (\ref{acceleration}) becomes
\be
\sqrt{\frac{\mu^2+\nu^2}{\al^2+\bb^2}}>\2
\ee
In the next section, when we discuss the dynamics involving a gas of
branes, we will see that, although the system does not reduce to a
single field dynamics, one can get power law inflation in a manner
similar to the case discussed above.

As a next step then, let us calculate the density fluctuations and
spectral tilt in this model. From (\ref{psi-evol}) and (\ref{a-evol})
one can compute $\dot{\psi'}$ and $H$ and hence the amplitude of
density fluctuations
\bea
\de_H &=& \frac{H^2}{2\pi\dot{\psi'}}=\frac{(\mu'+\al')H}{3\pi M_p} \\
&=&\frac{(\mu'+\al')}{3\pi M_p}\sqrt{\frac{V_0(\mu'+\al')e^{-2\al'\psi'}}{3\mu' M_p^2}} \nonumber
\eea
or 
\be
\de_H^2\sim a^{-\frac{3(1+\om)\al'}{\mu'+\al'}} \, , 
\ee
where the right hand side is evaluated at the time when the length
scale of the fluctuation being considered is exiting the Hubble radius
(see e.g.  \cite{MFB,LLbook} for reviews).  Therefore, the spectral
tilt $\eta$ is given by
\be
\eta\approx 1+\frac{1}{a}\frac{d\ln\de_H^2}{da}=1-\frac{3(1+\om)}{1+\mu'/\al'} 
\label{eta}
\ee
It is clear from (\ref{eta}) that in order to explain the observed
spectral tilt of the CMB spectrum (see e.g. \cite{WMAP}), $\eta>0.94$,
one needs a very large ratio $\mu'/\al'\sim {\cal O}(10)$ which is
difficult to achieve from string theory. Hence, it seems likely that
one needs to supplement this inflationary mechanism with an additional
mechanism to generate the observed almost scale-invariant spectrum of
density perturbations.  We will comment on this later.

{\bf Two Field Dynamics:} To solve the evolution equations
(\ref{phi}-\ref{Hubble}) for the two field case we choose the usual
ansatz
\be
a=a_0\left(\frac{t}{t_0}\right)^m\ ;e^{\psi}=e^{\psi_0}\left(\frac{t}{t_0}\right)^n\ ;e^{\phi} =e^{\phi_0}\left(\frac{t}{t_0}\right)^p
\label{e-ansatz}
\ee
In order that all the terms in (\ref{phi}-\ref{Hubble}) have the same
$t$ dependence, $t^{-2}$ to be specific, we get by looking at the
exponents
\be
\al'n+\bb'p=1=\frac{3m}{2}-\mu'n
\label{np}
\ee
{F}urther, one can substitute the potentials associated with $V_0$ and
$\rho_0$ from (\ref{phi},\ref{psi}) in (\ref{Hubble}) to obtain
another relation between the power law exponents:
\bea \label{m}
m^2 &=& \3\bigl[\2(n^2+p^2) \\
&+& (3m-1)(\frac{p}{2\bb'}(1+\frac{\al'}{\mu'})-\frac{n}{2\mu'})\bigr] \nonumber
\eea
One can thus solve (\ref{m}) and (\ref{np}) to obtain $m,n$ and $p$ in
terms of the exponents $\al',\bb'$ and $\mu'$ (see Appendix for
details). In particular one finds
\be
m=2\frac{3\mu'\al'+\al^{'2}+\bb^{'2}+2\mu^{'2}}{6\mu'\al'+3\al^{'2}+3\bb^{'2}+8\mu^{'2}\bb^{'2}}
\label{m-soln}
\ee
In order to find out whether one gets inflation, one has to check
whether $m > 1$ or not.

Intuitively, it is clear what needs to happen in order to have an
accelerated expansion: Along the $\psi'$ direction, due to the
coupling, one has a slowly evolving minimum due to the two opposing
exponential potentials as before. The field, though, can also roll
along the $\phi'$ direction, since $\bb'\neq 0$. However, if $\bb'$ is
sufficiently small (in other words if the two exponents corresponding
to the potential $V_0$ and $\rho_0$ are approximately collinear) then
the field rolls slowly also along the $\phi'$ direction and thus we
can have acceleration.

{F}inally, one has to check that the ansatz (\ref{e-ansatz}) gives
meaningful solutions, i.e. solutions exist for positive $V_0$ and
$\rho_0$. We show in the Appendix that this is indeed the case
provided $m>1/3$, which is obviously true for inflationary solutions.

\section{Supergravity and Brane Gases}

{\bf Supergravity and Effective Potentials:} Let us start with a
typical bosonic sector of a supergravity theory:
\bea\label{eq:sugra}
{\hat{S}} &=& \frac{1}{16\pi {\hat{G}}} \int d^{\Dh}x\sqrt{-g}\bigl[{\hat{R}}
- \p_{\mh}\phi\p^{\mh}\phi \\
&-& {1 \over 2}\sum_i e^{-2a_i\phi}{1 \over {n_i!}} F^2_{n_i} - V(\phi)\bigr] \nonumber
\eea
where hatted quantities denote the full higher ($\Dh$) dimensional
objects, $\phi$ is the dilaton field and the field strengths
$F_{n_i}$'s are $n_i$ forms with $i=1\dots M$. For generality, we have
also included a potential for the dilaton which may have both a
classical \cite{classical} or a quantum \cite{quantum} origin
depending on the specific supergravity/string theory model. In
general, the potential runs to either $\pm\infty$, leading to the much
studied issue of dilaton stabilization. For the purposes of this paper,
we do not concern ourselves with this issue and simply assume that
either it is stabilized after inflation or it is coupled very weakly
(if at all) to standard model particles and therefore even though the
dilaton slowly runs towards infinity, constraints coming from fifth
force experiments and variation of physical constants \cite{variation}
are satisfied. Indeed a ``least coupling mechanism'' has been proposed
\cite{least} to explain why the dilaton may couple very weakly to the
ordinary standard model particles. As a simple example of a running
potential we choose
\be V(\phi)=V_0e^{-2\de\phi} \ee 
We mention in passing that such
exponential potentials are found in several supergravity theories
\cite{classical}.

To obtain an effective four-dimensional theory one has to perform a
consistent dimensional reduction \cite{duff}. For a flux
compactification the only consistent ansatz (without involving
squashing) is given by \cite{exppot,flux}
\begin{equation}
\gh_{\mh\nh}=\left( \begin{array}{cc}
g_{mn}(x) & 0\\
0 &e^{2\psi(x)}\gs_{\ms\ns}(y)
\end{array} \right)
\label{eq:metric}
\ee
and
\be
F_{\ms_1\dots\ms_{\Ds}}\sim \e_{\ms_1\dots\ms_{\Ds}}
\ee
where $F$ has to be a $\Ds$-form, $\Ds$ being the number of extra
dimensions and we use the symbol ``$\circ$'' to indicate extra
dimensional quantities.  Once one solves the Bianchi identity and the
field equations for $F$, the $F^2$ term in the action (\ref{eq:sugra})
gives us a potential term for the scalars, $\phi$ and $\psi$. After
performing the usual dimensional reduction by integrating the extra
dimensions and applying conformal transformations
\be
\gh\ra e^{-\Ds\psi/(\Ds+2)}\gh
\ee
followed by 
\be
\psi\ra \sqrt{\frac{\Ds(\Ds+2)}{2}}\psi
\label{conf-psi}
\ee
to go to the Einstein frame in four dimensions, one finds that
\be
\int d^{\Dh}x\sqrt{-\gh}\frac{e^{-2a_i\phi}}{n_i!}F^2_{n_i}
\ee
becomes
\be 
\int d^4x\sqrt{-g}\cs^2e^{-2(a\phi+\ga' \psi)}
\ee
with
\be
2\ga'\equiv 3\sqrt{\frac{2\Ds}{\Ds+2}}
\ee
and $a$ is the same exponent that appears in the dilaton coupling to
the $\Ds$-form in (\ref{eq:sugra}), its value depends on the specific
supergravity model. Also, $\cs$ is a constant determining the strength
of the flux background.

Let us now compute the four dimensional effective potential coming
from the dilaton potential $V(\phi)$. We observe that dimensional
reduction followed by conformal transformation (\ref{conf-psi}) gives
a $\psi$ dependence which is in fact the same as one gets from a
higher dimensional cosmological constant. The four dimensional
potential reads
\be
V(\phi,\psi)=V_0e^{-2(\de\phi+\ga\psi)}
\label{dilaton}
\ee
with 
\be
2\ga=\sqrt{\frac{2\Ds}{\Ds+2}}
\ee
The full four dimensional action then reads
\bea\label{eq:4dgravity} S &=& \2\int
d^4x\sqrt{-g}\bigl[R-\p_{m}\phi\p^{m}\phi-\p_{m}\psi\p^{m}\psi
\nonumber \\ &-& 2(V_0'e^{-2(a\phi+\ga'
\psi)}+V_0e^{-2(\de\phi+\ga\psi)})\bigr] 
\eea 
where as before we have set $M_p=1$, and redefined $c_0$ in terms of
$V_0'$.

As a specific example, let us consider type IIA string theory. In the
string frame the relevant bosonic part of the action reads
\bea
\hat{S}_{II} &=& \int d^{10}x\sqrt{-g}e^{-2\phi}\bigl[\hat{R}+4\p_{\mh}\phi\p^{\mh}\phi-V(\phi)\bigr] \nonumber \\
&+& {1 \over 2}\frac{1}{4!}F^2_{4}+\dots
\label{typeII}
\eea 
where we have only included the 4-form whose dual is a 6-form field
strength and hence can be consistently turned on as there are 6 extra
dimensions.  The ellipsis involve other form and fermionic fields
which are set to zero as usual. To be general we have also added a
potential for the dilaton which could result from quantum corrections.
Performing the well known conformal transformation
\be
\hat{g}_{\mh\nh}\ra e^{-4\phi/(\Ds+2)}\hat{g}_{\mh\nh} 
\ee
followed by 
\be
\phi\ra {\sqrt\frac{4}{\Ds+2}}\phi
\label{conf-phi}
\ee
one recovers an action of the form (\ref{eq:sugra}) in the
10-dimensional Einstein frame with $2a_4=-1/\sqrt{2}$ corresponding to
the four form coupling exponent, or in terms of the 6-form dual,
$2a_6=1/\sqrt{2}$:
\be
\frac{1}{4!}e^{\phi}F^2_{4}\leftrightarrow\frac{1}{6!}e^{-\phi}F^2_{6} \, .
\ee 

The dimensionally reduced 4-dimensional action then looks like
(\ref{eq:4dgravity}) with
\be
2a_6=\frac{1}{\sqrt{2}}\mx{ and }2\ga'=3\sqrt{\frac{3}{2}}
\ee
where $\ga'$ and $a_6$ can be identified with $\al$ and $\bb$ of 
section 2 respectively.

Instead of looking at a flux background one may also study a potential
of the form (\ref{dilaton}) coming from a higher dimensional dilatonic
potential where substituting $\Ds=6$ we get
\be
2\ga=\sqrt{\frac{3}{2}}
\label{10-gamma}
\ee
and $\de$ remains a free parameter.

{\bf Brane Stress Energy:} Let us now consider a gas of branes wraping
all the compact internal dimensions and hence these are
$\Ds$-branes. The action for such a gas is given by
\be
S_{\mt{brane}}=\int d^{\Dh}x \sqrt{-\gh}\rho_0e^{-2\nu\phi}e^{-3\al}=\int d^{\Dh}x \sqrt{-\gh}\hat{\rho}
\ee
where $e^{\al}=a/a_0$ denotes the usual scale factor of our observable universe.
The exponential involving the dilaton in the last term originates from
the dilaton coupling present in the brane action in the string frame
which depends on several factors like the nature of the brane/string
(whether it is fundamental or solitonic etc. \cite{coupling}), the
dimensionality of the supergravity model, etc.  The second exponential
corresponds to the well known fact that the brane energy density
redshifts as non-relativistic dust along the transverse directions,
which in this case are the three large spatial dimensions.

In order to obtain the brane energy density in the 4-dimensional
Einstein frame one has to perform the conformal rescalings
(\ref{conf-psi}), so that in terms of the rescaled $\psi$ one has
\be
S_{\mt{brane}}=\int d^4x \sqrt{-g}\rho_0e^{2(\mu\psi+\nu\phi)}\left(\frac{a}{a_0}\right)^{-3}
\label{brane}
\ee
with
\be
2\mu=\sqrt{\frac{\Ds}{2(\Ds+2)}} \, ,
\label{mu}
\ee
or
\bea
S_{\mt{brane}} &=&\int d^4x \sqrt{-g}\rho_0e^{2\mu'\psi'}\left(\frac{a}{a_0}\right)^{-3} \nonumber \\
&=& \int d^4x \sqrt{-g}\ti{\rho}
\eea
where $\psi'$ is now the linear combination of the dilaton and the
radion as defined earlier in (\ref{linear}). $\ti{\rho}$ corresponds
to the observed four dimensional energy density of the wrapped branes.

Let us focus our attention now onto the 10 dimensional superstring
theories.  The DBI action for a $p$-brane in string frame is given by
\cite{leigh}
\be
S_{DBI}=T_p\int d^{p+1}\s e^{-2\phi}\sqrt{-\ga}
\label{dbi}
\ee
where $\ga$ is the induced metric on the world volume of the brane
parameterized by $\s$'s. (\ref{dbi}) leads to an action for a gas of
6-branes wrapping all the compact extra six directions, which looks
like
\be
S_{\mt{brane}}=\int d^{10}x \sqrt{-\gh}\rho_0e^{-2\phi}e^{-3\al}
\ee
in the string frame. Conformal transformation of $\gh$
(\ref{conf-phi}) and subsequent rescaling of $\phi$ then gives us the
action in Einstein frame (\ref{brane}) with
\be
2\nu=\frac{\Ds}{2}\sqrt{\frac{1}{\Ds+2}}=\frac{3}{\sqrt{8}}
\ee
while substituting $\Ds=6$ in (\ref{mu}) gives us the exponent $\mu$
\be
2\mu=\sqrt{\frac{3}{8}} \, .
\ee
Moreover, one finds in this case
\be
\psi'= \cos(\pi/3) \psi+\sin(\pi/3)\phi\mx{ and }\mu'=\sqrt{\frac{3}{8}} \, .
\label{10-mu}
\ee

{\bf Inflation from 10 Dimensional Universe:} We have finally
accumulated all the objects neccessary to understand whether or not
one can indeed obtain a phase of acceleration in the early universe
with branes in conjunction with string theory potentials for modulii
fields and the dilaton. Let us first look at the case when no flux is
turned on ($V_0'=0$ in (\ref{eq:4dgravity})) but rather we have a
dilatonic potential of the form (\ref{dilaton}). In this case we have
one free parameter, namely $\de$, and whether one has acceleration or
not depends on it. First, realize that in order to end inflation we
want
\be
\al_{eff}\equiv\sqrt{\al^{'2}+\bb^{'2}}>1
\label{tracking}
\ee
This is because, once the branes have decayed, the scalars evolve as
if under the influence of an exponential potential with an effective
exponent $\al_{eff}$ \cite{flux}. Therefore, in order for the scalars
to track radiation \cite{copeland} after the end of inflation we
require (\ref{tracking}). (\ref{tracking}) implies a bound
$\de>\sqrt{5/8}$.  Now, plugging in all the relevant exponents in the
expression for $m$ in (\ref{m-soln}) one finds that it is possible to
have an inflationary paradigm provided
\be
\sqrt{\frac{5}{8}}<\de<\sqrt{\frac{9}{8}} \, ,
\label{range}
\ee
For these values of $\de$ one finds
\be
1<m<1.1
\ee
in other words, when $\de$ is close to 1, and this seems reasonable
from the string theory point of view.

Next, let us look at the potential coming from the 4-form flux.
Substituting the exponents $a_6$ and $\ga'$ in (\ref{m-soln}) one
finds
\be
m<1
\ee
Thus, with just a 4-form flux and branes one cannot get acceleration.
This situation however can change if a seperate mechanism is available
to stabilize a particular linear combination of the dilaton and the
radion.

\section{Graceful Exit and Reheating}

Branes in string theory can also be interpreted as solitonic solutions
in the corresponding low energy supergravity theory. One subclass of
such branes are the unstable branes of string theory
(e.g. even-dimensional branes in Type IIB string theory). Provided
that the tachyon which describes the decay of these branes interacts
with fields which are in thermal equilibrium in the early universe,
these branes could be stabilized at early times - like the embedded
Z-string in the standard electroweak theory \cite{Nag} - and thus
trigger inflation as described above. The phase of inflation would
last until the fields which mediate the interaction, e.g. the photon
in the case of the standard model Z-string, fall out of
equilibrium. At that time, the tachyonic instability could set in. The
energy density stored in the unstable branes would lead to
reheating. In the example below, the role of the bulk modes is played
by the Kaluza-Klein modes.

To be specific, we consider a subclass of brane-like solutions of the
supergravity equations known as black branes \cite{bb} which appear to
us (i.e. in the effective four dimensional world) as black
holes. Stability of such black branes (both charged and uncharged) has
been discussed in detail in \cite{stab}, and in particular it was
realized that when the Kaluza-Klein modes of the gravitational
supermultiplet are lighter than the tension of the branes, these modes
become unstable (Gregory-Laflamme instability)
\cite{stab}\footnote{Stability of branes depends on various factors
like the charge and the dilatonic coupling of the branes
\cite{dilaton}, but for simplicity we will consider uncharged branes
and assume that the dilatonic couplings do not run.}.  Since in our
model the volume of the internal manifold slowly grows, the
Gregory-Laflamme instability thus provides us with a natural mechanism
to end inflation. Assume that the mass scale associated with the
tension of the branes is given by
\be
T=10^{-B}M_p
\ee
The mass of the Kaluza-Klein modes are, on the other hand, given by
\be
M_K=e^{-\ze\psi}M_p\mx{ with }\ze=\sqrt{\frac{\Ds+2}{2\Ds}}
\ee
The branes become unstable when $M_K\sim T$. In terms of the rotated
basis this happens when
\bea
10^{B} &=& e^{\ze(\psi'\cos\te-\phi'\sin\te)} \\
&=& e^{\ze(\psi'_0\cos\te-\phi'_0\sin\te)}\left(\frac{t}{t_0}\right)^{\ze(n\cos\te-p\sin\te)} \nonumber \\
&=& e^{-\ze\psi_0}\left(\frac{a}{a_0}\right)^{\ze(n\cos\te-p\sin\te)/m} \nonumber \\
&\equiv& N_0e^{{\cal N}\ze(n\cos\te-p\sin\te)/m} \nonumber
\eea
where ${\cal N}$ is the number of e-foldings. In the spirit of string
gas cosmology \cite{BV} we assume that, initially, the internal volume
is of Planck size, i.e. $\psi\approx 0$. This implies $N_0 \sim {\cal
O}(1)$. One can then estimate the number ${\cal N}$ of e-foldings as
\be
{\cal N}=\frac{2.3Bm}{\ze(n\cos\te-p\sin\te)}
\ee
For a range of the relevant parameters one can indeed obtain a sufficient 
number of efoldings. For example, for a ten-dimensional model, substituting 
$\Ds=6$ and $\te=\pi/3$ one finds
\be
{\cal N}=\frac{2.3Bm}{0.41n-0.71p}
\ee
For the exponent calculated for a gas of six branes (\ref{10-mu}) with
a dilatonic potential (\ref{dilaton},\ref{10-gamma}), $\de$ within the
range (\ref{range}), one finds that for a moderate hierarchy, $B\sim
5-7$ we get around 50-60 efoldings which is neccessary to solve the
flatness and horizon problems.

After the black-branes have decayed, the field starts rolling along the 
exponential potential of the dilaton and the radion. At late times, there 
are two possible scenarios. If the potential is indeed a pure exponential 
as we considered in our model, then the fields would continue to roll 
tracking first the radiation and later on the matter energy density 
(since $\sqrt{\al^2+\bb^2}>1$). One can try to connect this model to a late 
time coupled quintessence regime as discussed in \cite{BM}. However, 
the phenomenological viability of this scenario requires a mechanism which 
ensures that the standard model particles are only very weakly coupled to 
radion and the dilaton - otherwise one has conflicts with fifth force 
experiments. Indeed, in \cite{least} such a least coupling principle was 
proposed for the coupling of the dilaton to standard model particles. We here 
do not further speculate on the details of such mechanisms. The other 
possibility could be if the radion-dilaton potential has a minimum where the 
fields could be stabilized after the end of inflation. This possibility
might be realized if, for example, the potential is really a sum of 
exponentials. This occurs quite commonly in supergravity reductions 
\cite{exppot,flux}. In this case, of course, the moduli are stabilized and 
one does not need to worry about the fifth force constraints 
(as long as the potential is sufficiently curved at the minimum).

\section{Discussion and Conclusions}

We have presented a mechanism for obtaining inflation in the context
of string gas cosmology. We start with the conventional scenario of
string gas cosmology: the universe starts out hot and small (string
scale), with all spatial dimensions of comparable scale. As discussed
in \cite{BV,ABE,BEK}, the fundamental string winding modes can
disappear in at most three spatial dimensions, thus leading to an
explanation of why the spatial dimensions predicted by string theory
but not seen experimentally are confined. Radion stabilization is a
natural result of the string winding and momentum modes about the
extra spatial dimensions \cite{Patil}. 

At the later times studied in
this paper, we have focused on the effective four-dimensional field
theory coming from dimensionally reducing the Lagrangian of string gas
cosmology. Assuming the existence of an exponential potential for the
dilaton and the radion at these later times, we have shown that the
coupling of branes winding the extra dimensions couple to the radion
and dilaton and can lead to a period of power-law inflation.  This
inflationary period is generated by the combined radion-dilaton system
tracking its time-dependent potential minimum position. The
time-dependence of the effective potential is induced by the coupling
of the brane gas to the two fields. A graceful exit from inflation is
obtained by making use of a gas of unstable branes, branes which are
stabilized at early times via their interactions with the Kaluza-Klein
modes of the fields of the bulk gravitational supermultiplet. The
decay of these branes then leads to reheating after inflation.

For the specific branes we considered we obtained a power law exponent
$m$ which is too small to be consistent with the observational limits
on the tilt in the power spectrum of density fluctuations. In
addition, a hierarchy between the Planck scale and the tension of the
branes is required in order to obtain a sufficient number of
e-foldings of inflation. Both of these problems can be solved provided
the brane coupling exponents are larger than what we obtained for
6-branes in the context of 10 dimensional supergravity/string theory.
This can happen in several ways. As noted in \cite{quantum},
quantum stingy loop corrections can change several coefficients in the
effective dilatonic action which in turn will change the exponents
(For example, if the coefficient in front of the kinetic term for the
dilaton changes, one has to rescale the dilaton field which in turn
changes its coupling exponent to the brane.). 

One could also consider
various types of branes with different dimensionalities, perhaps in
the context of lower dimensional supergravity models. The various
exponents again are expected to be different, as is evident from the general analysis we
performed. Finally, it is not neccessary to restrict oneself only to
radion-dilaton systems. For example, one could consider squashed
configurations where the branes couple to the volume as well as the
shape moduli (the dilaton would be stabilized by some other
mechanism). As shown in \cite{flux}, for such configurations it is
possible to consistently turn on other lower form fluxes (not just the
6-form as considered here) which may come with the right exponents to
realize an accelerated regime in conjunction with the branes. We leave
an exploration of all these possibilities for the future.
\section{Appendix: Two field Solution}
We want to solve cosmological evolution equations of the form
(\ref{psi}-\ref{Hubble}): 
\be \ddot{\phi}+3H\dot{\phi}=2\bb
V_0e^{-2(\al\psi+\bb\phi)}
\label{aphi}
\ee
\be
\ddot{\psi}+3H\dot{\psi}=2\al V_0e^{-2(\al\psi+\bb \phi)}-2\mu\rho e^{2\mu \psi}
\label{apsi}
\ee
\begin{eqnarray}
\label{aHubble}
H^2 =\3\left( \frac{\dot{\phi}^2}{2}+\frac{\dot{\psi}^2}{2}+V_0e^{-2(\al\psi+\bb \phi)}+\rho e^{2\mu \psi}\right)\,.
\end{eqnarray}
We choose the ansatz as in (\ref{e-ansatz})
\be
a=a_0\left(\frac{t}{t_0}\right)^m\ ;e^{\psi}=e^{\psi_0}\left(\frac{t}{t_0}\right)^n\ ;e^{\phi} =e^{\phi_0}\left(\frac{t}{t_0}\right)^p
\label{a-ansatz}
\ee
In order that all the terms in (\ref{apsi},\ref{aHubble}) have the
same $t$ dependence, $t^{-2}$, to be specific, we get by looking at
the exponents 
\be \al n+\bb p=1=\frac{3m}{2}-\mu n\label{a-np}
\ee
Further, one can substitute the potentials associated with $V_0$ and
$\rho_0$ from (\ref{apsi}) in terms of the field derivatives in
(\ref{aHubble}) to obtain
\bea H^2 &=&\3\left(
\frac{\dot{\phi}^2}{2}+\frac{\dot{\psi}^2}{2}-\frac{1}{2\mu}(\ddot{\psi}+3H\dot{\psi})
\right.\nonumber\\
&-&\left.\frac{1}{2\bb}(1+\al/\mu)(\ddot{\phi}+3H\dot{\phi})\right)
\eea 
Substituting the ansatz (\ref{a-ansatz}) in the above equation one
finds another relation between the power law exponents
\be
m^2=\3\left[\2(n^2+p^2)+(3m-1)(\frac{p}{2\bb}(1+\frac{\al}{\mu})-\frac{n}{2\mu})\right]
\label{a-m}
\ee
One can thus solve (\ref{a-m}) and (\ref{a-np}) to obtain $m,n$ and
$p$ in terms of the exponents $\al,\bb$ and $\mu$: 
\be
m=2\frac{3\mu\al+\al^{2}+\bb^{2}+2\mu^{2}}{6\mu\al+3\al^{2}+3\bb^{2}+8\mu^{2}\bb^{2}}
\label{am-soln}
\ee
\be
n=\frac{3\al+6\mu-8\bb^{2}\mu}{6\mu\al+3\al^{2}+3\bb^{2}+8\mu^{2}\bb^{2}}
\label{n-soln}
\ee
\be
p=\frac{\bb(8\mu^2+3+8\mu\al)}{6\mu\al+3\al^{2}+3\bb^{2}+8\mu^{2}\bb^{2}}
\label{p-soln}
\ee
Although the $t$ dependence now cancels in all the evolution equations
by virtue of (\ref{am-soln}-\ref{p-soln}), we are still left with
matching the coefficients in the two equations (\ref{aphi}) and
(\ref{apsi}). These equations essentially determine the other unknown
parameters $\phi_0$ and $\psi_0$ in terms of $V_0$ and $\rho_0$ or
vice-versa. In fact to have a consistent solution one should check
whether these equations can be satisfied for positive values of $V_0$
and $\rho_0$ as those are the physical scenarios we are interested
in. After some algebra one finds
\be
V_0=e^{-2(\al\psi_0+\bb\phi_0)}\frac{p(3m-1)}{2\bb}
\label{Vnot}
\ee
and
\bea
\rho_0=e^{-2\mu \psi}\left(\frac{3m-1}{2\mu}\right)\left(\frac{\al}{\bb}p-n\right)\nonumber\\
=e^{-2\mu \psi}(3m-1)\left(\frac{4(\al^2+\bb^2)-3+4\mu\al}{6\mu\al+3\al^{2}+3\bb^{2}+8\mu^{2}\bb^{2}}\right)
\label{rhonot}
\eea
From (\ref{p-soln}) it is clear that the ratio $p/\bb$ is
positive. Also, for the solutions we are looking at, $m>1>1/3$, and
thus the right hand side of (\ref{Vnot}) is positive implying
$V_0>0$. Next let us look at the right hand side of (\ref{rhonot}).

In order for the potential to track radiation after inflation we
demanded $\al^2+\bb^2>1$ and thus it is clear that $\rho_0$ is
positive too. We indeed have consistent attractor solutions.

\begin{acknowledgements}
RB is supported in part (at McGill) by an NSERC Discovery grant and
(at Brown) by the US Department of Energy under Contract
DE-FG02-91ER40688, TASK~A. The work of TB is supported by NSERC Grant
No.\ 204540. DE is supported in part by NSF-PHY-0094122 and funds from
Syracuse University.
\end{acknowledgements}


\end{document}